\documentclass[aps,reprint]{revtex4-1}

\usepackage[english]{babel}
\usepackage{amsfonts}
\usepackage{amsmath}
\usepackage{graphicx}
\usepackage{mathtools}
\usepackage[caption=false]{subfig}
\usepackage{multirow}

\usepackage{soul}
\usepackage{color}

\newcommand{\C}[1]{$^{\circ}$C}

\begin{document}

\title{Prediction of crystallized phases of amorphous Ta$_2$O$_5$-based mixed oxide thin films using density functional theory calculations}

\author{Mariana A. Fazio}
\email{Mariana.Fazio@colostate.edu}
\affiliation{Department of Electrical and Computer Engineering, Colorado State University, Fort Collins, CO, USA}

\author{Le Yang}
\affiliation{Department of Chemistry, Colorado State University, Fort Collins, CO, USA}

\author{Carmen S. Menoni}
\email{Carmen.Menoni@colostate.edu}
\affiliation{Department of Electrical and Computer Engineering and Department of Chemistry, Colorado State University, Fort Collins, CO, USA}
\date{\today}


\begin{abstract}
	
The genomics approach to materials, heralded by increasingly accurate density functional theory (DFT) calculations conducted on thousands of crystalline compounds, has led to accelerated material discovery and property predictions. However, so far amorphous materials have been largely excluded from this as these systems are notoriously difficult to simulate. Here we study amorphous Ta$_2$O$_5$ thin films mixed with Al$_2$O$_3$, SiO$_2$, Sc$_2$O$_3$, TiO$_2$, ZnO, ZrO$_2$, Nb$_2$O$_5$ and HfO$_2$ to identify their crystalline structure upon post-deposition annealing in air both experimentally and with simulations. Using the Materials Project open database, phase diagrams based on DFT calculations are constructed for the mixed oxide systems and the annealing process is evaluated via grand potential diagrams with varying oxygen chemical potential. Despite employing calculations based on crystalline bulk materials, the predictions agree well with the experimentally observed crystallized phases of the amorphous thin films. Only in two cases the database leads to incorrect predictions: in TiO$_2$-doped Ta$_2$O$_5$ because it does not contain a ternary compound found experimentally, and in Sc$_2$O$_3$-doped Ta$_2$O$_5$ because DFT overestimates the formation enthalpy difference between Sc$_2$O$_3$ and Ta$_2$O$_5$ and thus does not reproduce observed oxygen competition effects. In the absence of ternary phases, the dopant acts as an amorphizer agent increasing the thermal stability of Ta$_2$O$_5$. The least efficient amorphizer agent is found to be Nb$_2$O$_5$, for which the cation has similar chemical properties to those of Ta in Ta$_2$O$_5$. These results show that DFT calculations can be applied for the prediction of crystallized structures of annealed amorphous materials. This could pave the way for accelerated \textit{in silico} material discovery and property predictions using the powerful genomic approach for amorphous oxide coatings employed in a wide range of applications such as optical coatings, energy storage and electronic devices.

\end{abstract}

\pacs{}

\maketitle

\section{Introduction}

Amorphous mixed oxide thin films have found their way into numerous applications ranging from high-k dielectrics to gas sensors to optical coatings with tunable properties \cite{stenzel2011mixed, choi2011development,zakrzewska2001mixed}. They constitute particularly interesting systems, consisting of a mixture of two traditionally well-known oxides, that produces a new material which in principle can be tailored to specific applications by varying the doping proportion. The introduction of the dopant oxide can serve different purposes. In the case of high-k dielectrics, some of the most promising materials such as HfO$_2$ or Ta$_2$O$_5$ are doped with other oxides to suppress their crystallization after being annealed at high temperatures because the presence of crystallites is correlated with an increase in the leakage current \cite{tewg2004suppression, choi2011development}. For optical interference coatings, oxide mixtures were investigated and optimized to achieve higher laser damage thresholds \cite{jupe2007mixed,jensen2010investigations}. One particular application of mixed oxide optical coatings is their use in the high-reflectance stacks of the gravitational-wave interferometers of Advanced LIGO \cite{abbott2016gw150914} and Advanced Virgo \cite{acernese2014advanced}. The current high index material is a mixture of Ta$_2$O$_5$ with around 20\% of TiO$_2$. The addition of TiO$_2$ led to a decrease in the internal friction of the coating compared to pure Ta$_2$O$_5$ \cite{harry2006titania,granata2016mechanical,granata2020amorphous,fazio2020structure}, which resulted in a significant increase of the sensitivity of the detectors giving way to remarkable astrophysical discoveries \cite{aasi2015advanced}. The mixing of two (or more) amorphous oxides can lead to the discovery of new materials with vastly improved performance. However, most research conducted on amorphous mixed oxide coatings has been an experimental process of trial-and-error and there is currently no theoretical framework to predict the properties of the amorphous mixture.

In the last decade, emerging computational methods have been conducive to accelerated materials discovery with the genomics approach being a valuable tool to predict phase formation and design new materials \cite{jain2016research}. The genomics approach to materials is heralded by increasingly accurate density functional theory (DFT) calculations that are conducted on thousands of compounds. The results of these calculations are readily available in open-access databases such as the Materials Project \cite{Jain2013} or the Open Quantum Materials Database \cite{saal2013materials}. An application that has yielded excellent results is phase stability studies for oxide systems via the construction of phase diagrams using DFT calculations. In particular, it has been used to predict reactions in cathode material for lithium batteries \cite{Ong2008} and for phase formation in steels under oxidation conditions \cite{chinnappan2014thermodynamic}. However, in all these approaches the materials under consideration were crystalline bulk or crystalline thin films for which DFT calculations yield accurate results. Amorphous systems have been largely excluded from the genomics approach because finding the right model to simulate their structures can be challenging and even more so if a mixture is under consideration, so there are no large databases of calculated structures available. But if DFT calculations proved to be useful to predict the crystallized phases of the amorphous systems, then these databases and the genomics approach could be used to gain insights into these technologically important materials. For instance, TiO$_2$-doped Ta$_2$O$_5$ amorphous films that exhibited low internal friction also coincidentally crystallized in a ternary compound after post-deposition annealing, indicating that the crystallized structure can be related to specific features of the amorphous phase \cite{fazio2020structure}. In order to assess crystalline phase stability for the amorphous materials, the effect of the annealing process needs to be incorporated. This can be done by using grand potential diagrams, which allows one to introduce thermodynamic processes into DFT \cite{Ong2008}. By studying the annealing process and evaluating the stability of phases in the crystalline systems, one might be able to predict the structure of crystallized mixed oxides which can provide information about the amorphous phase of the material. This could allow \textit{in silico} material design for a new range of applications involving amorphous materials such as those used in optical coatings, energy storage and electronic devices.



In this paper, we carried out an extensive study of Ta$_2$O$_5$ amorphous thin films doped with Al$_2$O$_3$, SiO$_2$, Sc$_2$O$_3$, TiO$_2$, ZnO, ZrO$_2$, Nb$_2$O$_5$ and HfO$_2$ grown by reactive sputtering. The influence of the dopant and post-deposition annealing on the film properties and structure was characterized by grazing-incidence x-ray diffraction and x-ray photoelectron spectroscopy. Using the Materials Project database, phase diagrams based on DFT calculations were constructed for all mixed oxide systems and the annealing process was evaluated via grand potential diagrams with varying oxygen chemical potential. Despite employing DFT calculations that correspond to crystalline materials, the predictions agree well with the experimentally obtained crystallized phases. Only in two cases the database leads to incorrect predictions: in TiO$_2$-doped Ta$_2$O$_5$ because it does not contain a ternary compound found experimentally, and in Sc$_2$O$_3$-doped Ta$_2$O$_5$ because DFT overestimates the formation enthalpy difference between Sc$_2$O$_3$ and Ta$_2$O$_5$ and thus does not reproduce observed oxygen competition effects. We find that predicted ternary compounds can be stabilized if their cation ratios are similar to or lower than the cation ratio of the film. When no ternary phases are present, the dopant acts as an amorphizer agent increasing the thermal stability of Ta$_2$O$_5$. The least efficient amorphizer agent is found to be Nb$_2$O$_5$, for which the cation has similar chemical properties to those of Ta in Ta$_2$O$_5$. The results of this study pave the way for applying DFT to the prediction of crystallized structures of annealed amorphous coatings, identifying suitable dopants to increase thermal stability and tailoring processing conditions for the production of ternary oxide films.





\section{Experimental} \label{sec:experimental}

The films were deposited by reactive ion beam sputtering employing the Laboratory Alloy and Nanolayer System (LANS) manufactured by 4Wave, Inc \cite{zhurin2000biased}. Experimental details can be found in \cite{fazio2020growth}. In this work, metallic targets of Ta, Al, Si, Sc, Ti, Zn, Zr, Nb and Hf of 99.99\% purity were employed. The pulse period was set to 100 $\mu$s and the oxygen flow to 12 sccm. Deposition conditions for all evaluated films are presented in Table \ref{table:dep-parameters}. Coatings were grown on 25.4 mm diameter and 6.35 mm thick ultraviolet grade fused silica substrates and on (111) Si wafer substrates. After deposition, the films were annealed in air until crystallization was reached using a Fisher Scientific Isotemp programmable muffle. A heating rate of 100\C{} per hour was employed and the samples were soaked for 10 hours at 300\C{}, 500\C{}, 600\C{}, 700\C{}, 800\C{} and 900\C{}.

%

\begin{table*}[!ht]
	\setlength\tabcolsep{5pt}
	\begin{center}
		\begin{tabular}{ccccc}
			\hline
			Film & Targets & Pulse width ($\mu$s) & Deposition rate (nm/s) & Dopant cation ratio\\
			\hline
			I & Ta & 2 & 0.0212 $\pm$ 0.0001 &  - \\
			II & Al - Ta & 51 - 2 & 0.0271 $\pm$ 0.0001 & 0.17 $\pm$ 0.01\\
			III &  Si - Ta & 72 - 2 & 0.0246 $\pm$ 0.0001 & 0.26 $\pm$ 0.01\\
			IV & Sc - Ta & 45 - 2 & 0.0287 $\pm$ 0.0001 & 0.105 $\pm$ 0.007\\
			V & Ti - Ta & 2 - 53 & 0.01603 $\pm$ 0.00005 & 0.27 $\pm$ 0.04\\
			VI & Zn - Ta & 56 - 2 & 0.0254 $\pm$ 0.0001 & 0.20 $\pm$ 0.01\\
			VII & Zr - Ta & 54 - 2 & 0.0294 $\pm$ 0.0001 & 0.23 $\pm$ 0.01\\
			VIII & Nb - Ta & 68 - 2 & 0.0261 $\pm$ 0.0001 & 0.12 $\pm$ 0.01\\
			IX & Hf - Ta & 49 - 2 & 0.0295 $\pm$ 0.0001 & 0.23 $\pm$ 0.02\\
			\hline
		\end{tabular}
	\end{center}
	\caption{Deposition conditions for the coatings in this study along with cation ratios determined from XPS atomic concentrations.}
	\label{table:dep-parameters}
\end{table*}

The films as deposited and after each annealing step were characterized by grazing incidence x-ray diffraction (GIXRD) and x-ray photoelectron spectroscopy (XPS). A Bruker D8 Discover Series I diffractometer with a Cu K$\alpha$ source was used for GIXRD with an incident angle of $0.5^{\circ}$ and $2\theta$ between $10^{\circ}$ and $80^{\circ}$. For XPS measurements a Physical Electronics PE 5800 ESCA/ASE system with a monochromatic Al K$\alpha$ x-ray source was employed. The photoelectron take-off angle was set at $45^{\circ}$. The instrument base pressure was around $1 \times 10^{-9}$ Torr and a charge neutralizer with a current of 10 $\mu$A was used for all measurements. The binding energy scale was calibrated based on the position of the adventitious carbon. From the atomic concentrations measured by XPS, the cation ratios (defined as M/(M +Ta) with M the dopant atomic concentration and Ta the tantalum atomic concentration) were determined and are shown in Table \ref{table:dep-parameters} ranging from 0.17 to 0.27.

DFT-based ternary phase diagrams were constructed for all mixed oxides using the PDApp \cite{Ong2008,Jain2011a} of the Materials Project (MP) database \cite{Jain2013}. These diagrams are constructed at 0 K for an isothermal, isobaric, closed system with the relevant thermodynamic potential being the Gibbs free energy which is equivalent to the internal energy in this case. The internal energy of all the relevant phases calculated using DFT is extracted from the MP database, which uses the generalized gradient approximation (GGA) and the GGA+U extension when appropriate. The errors associated with these calculations will be discussed in the Results section. However, as we want to study the effect of annealing on the stability of the phases, the system of interest is not completely closed but rather it is an isothermal, isobaric system open only with respect to oxygen. In this case, the oxygen grand potential is the appropriate thermodynamic potential and therefore grand potential diagrams were also constructed using PDApp. The effect of both the temperature $T$ and the oxygen partial pressure $p_{O_2}$ is fully captured in the oxygen chemical potential ($\mu_{O_2}$) that is defined as \cite{Ong2010}:

\begin{equation*}
\mu_{O_2} (T,p_{O_2}) = E_{O_2} + kT - T S_{O_2}(T,p_{0}) + kT ln(p_{O_2}/p_0)
\end{equation*}

with $E_{O_2}$ the energy of the most stable O$_2$ compound in the Materials Project database, $k$ the Boltzmann constant and $S_{O_2}(T,p_{0})$ the entropy at a temperature $T$ and a reference partial pressure $p_0$. Taking the reference partial pressure to be the atmospheric pressure $p_0 = 0.1 MPa$, the tabulated values for $S_{O_2}(T,p_{0})$ can be found in \cite{rumble2019crc}. In the case of this study, $\mu_{O_2}$ ranges from -5.6 eV to -8 eV, with increasing annealing temperature leading to a more negative oxygen chemical potential and thus a more reducing environment. The grand potential diagrams were evaluated for $\mu_{O_2}$ in the range of interest.
 

\section{Results} \label{sec:results}

\begin{table}[!ht]
	\setlength\tabcolsep{5.5pt} 
	\begin{tabular}{ccc}
		\hline
		\multirow{2}{*}{Oxide} & \multicolumn{2}{c}{Formation enthalpy (kJ/mol)}\\
		\cline{2-3}
		& Calculated at 0K & Experimental at 298K\\
		\hline
		Ta$_2$O$_5$ & -2264 & -2046.0 \\
		Sc$_2$O$_3$ & -1919 & -1908.8 \\
		Nb$_2$O$_5$ & -2057 & -1899.5 \\
		Al$_2$O$_3$ & -1657 & -1675.7 \\
		HfO$_2$ & -1167 & -1144.7 \\
		ZrO$_2$ & -1107 & -1100.6 \\
		TiO$_2$ (rutile) & -1006 & -944.0 \\			
		SiO$_2$ & -949 & -910.7 \\
		ZnO & -347 & -350.5 \\
		\hline
	\end{tabular}
	\caption{Comparison of formation enthalpy for all evaluated oxides: calculated from MP database and from experiments \cite{rumble2019crc}.}
	\label{table:thermo-prop}
\end{table}

Table \ref{table:thermo-prop} shows a comparison between the formation enthalpies of the oxides under consideration extracted from MP and the experimental values measured at 298 K. The calculated formation enthalpy at 0 K provided by MP is equivalent to the formation energy of the compound, which is the key parameter to construct the phase diagrams. There is a remarkably good agreement between the calculated values at 0 K and the experimental values at 298 K, which is in part why the DFT phase diagrams are also able to predict fairly well phase stability at room temperature for crystalline oxide systems \cite{chinnappan2014thermodynamic}.

\begin{figure*}[h!]
	\includegraphics[width=0.8\linewidth]{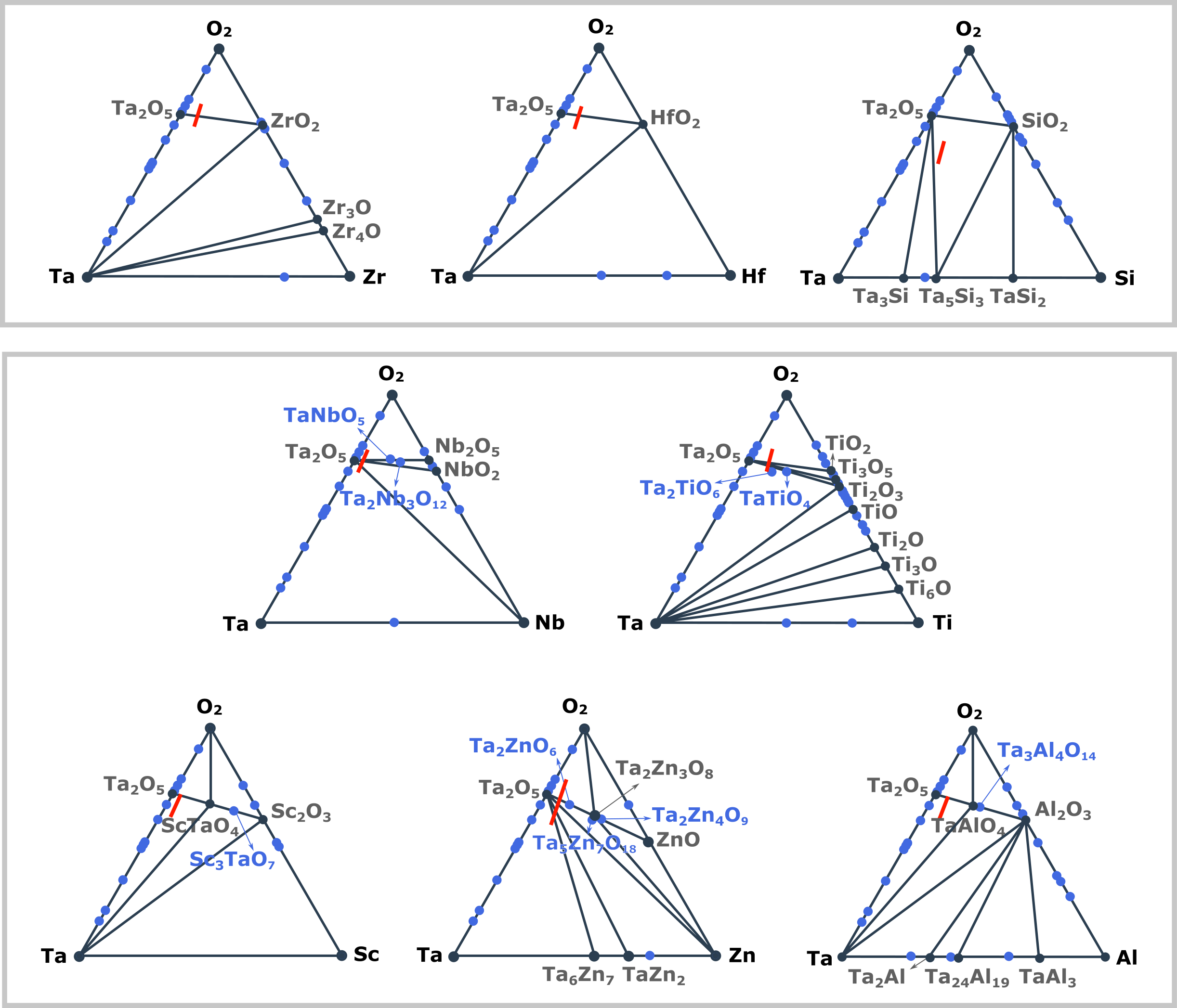}
	\caption{\label{fig:phase-diag} Phase diagrams at 0 K constructed with DFT calculations from the MP database for all the evaluated systems. The red line (compositional line) indicates the range of atomic concentrations obtained by XPS for each mixture as-deposited and after annealing up to the crystallization temperature.}
\end{figure*}

Based on the formation energies provided by MP, ternary phase diagrams were constructed using the PDApp for all evaluated oxide mixtures and are presented in figure \ref{fig:phase-diag}. These diagrams have a triangular shape with its vertices being the two metallic cations of the mixture and gaseous oxygen respectively. The solid lines are constructed by projecting the 3-dimensional energy convex hull into the compositional space and form Gibbs triangles. The nodes of the Gibbs triangles (indicated in black) represent the phases with the lowest energy, that is the stable phases of the system. In addition, other phases predicted to be unstable are also included as blue points. These unstable phases are regarded as so because they have an energy above hull higher than zero and thus will spontaneously decompose into compounds at the endpoints of their corresponding convex hulls. For each phase diagram, the red line indicates all compositions consistent with the atomic concentrations obtained by XPS for each mixture. The oxygen was varied $\pm$ 5\% to account for observed variations with the annealing temperature and only the oxygen not bonded with carbon species was considered \cite{payne2011x}.

Figure \ref{fig:phase-diag} is composed of two panels indicating the distinct effects of the dopant addition. The top panel includes the systems for which no ternary compounds are predicted, with all phases located along the sides of the triangle. This is the case for three systems: Zr-Ta-O, Hf-Ta-O and Si-Ta-O. Based on the compositional lines of the mixtures, one or several Gibbs triangles are intercepted with multiple predicted phases. For all these systems Ta$_2$O$_5$ is a stable phase along with the oxide formed by the dopant (ZrO$_2$, HfO$_2$ and SiO$_2$). Tantalum and other metallic alloys are also predicted for these systems. For the rest of the systems in the bottom panel of figure \ref{fig:phase-diag}, the dopant addition is predicted to induce the formation of ternary oxide phases, which are located in the center of the diagrams. In the case of Ti-Ta-O and Nb-Ta-O, the ternary phases are predicted to be unstable. For Nb-Ta-O, based on the elemental composition of the film, the compositional line intersects all Gibbs triangles while for Ti-Ta-O system, the compositional line can be found within two Gibbs triangles with one including the unstable ternary compounds. Lastly, for Sc-Ta-O, Zn-Ta-O and Al-Ta-O the phase diagrams predict both stable and unstable ternary phases. For the Sc-Ta-O system, based on the atomic concentrations of the film, the predicted stable phases are Ta, Ta$_2$O$_5$ and ScTaO$_4$. In the case of the Zn-Ta-O system, five Gibbs triangles are intercepted with the compositional line which results in several predicted phases including ternary compounds with various cation ratios. For the predicted ternary phases only Ta$_2$Zn$_3$O$_8$ is marked as stable while two other ternary compounds, Ta$_2$ZnO$_6$ and Ta$_2$Zn$_4$O$_9$, are marked as unstable. For Al-Ta-O the phase diagram shows that the compositional line for the film lies within a single Gibbs triangle with the predicted phases being Ta, Ta$_2$O$_5$ and TaAlO$_4$.

\onecolumngrid

\twocolumngrid

\begin{figure*}[h!]
	\includegraphics[width=0.8\linewidth]{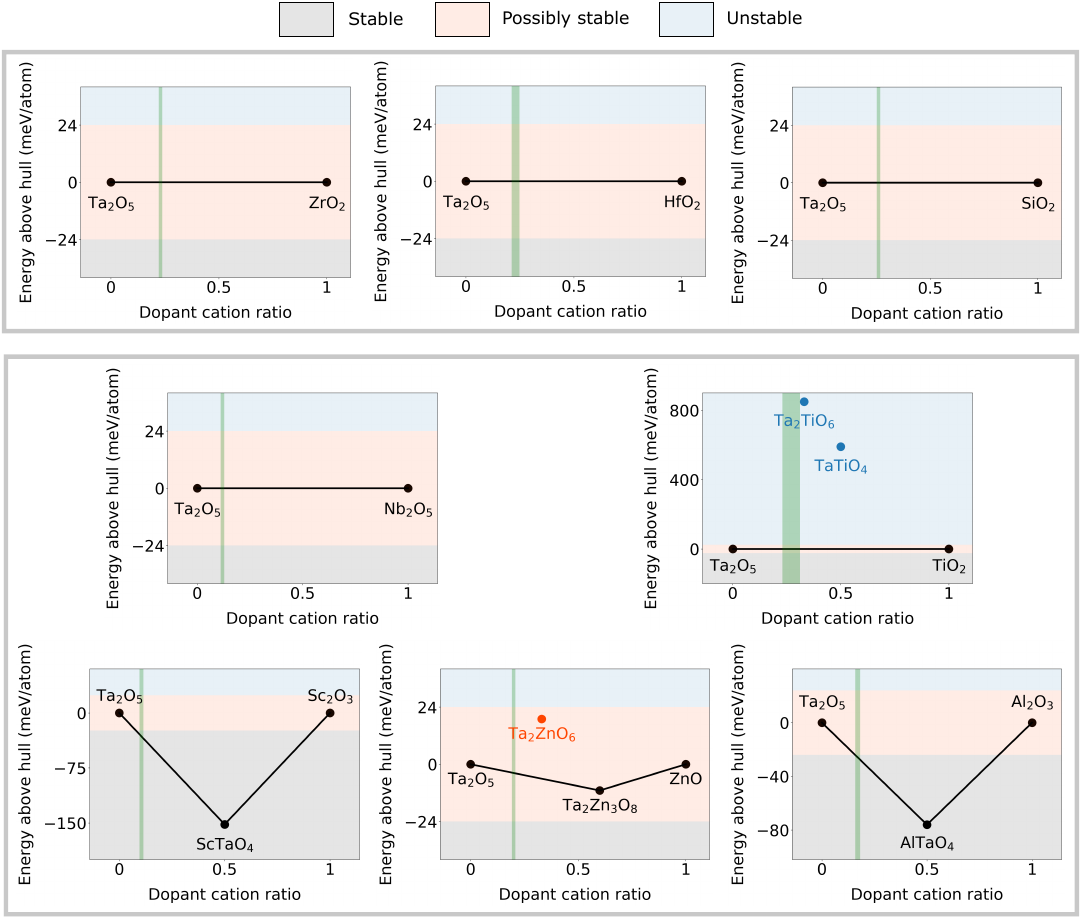}
	\caption{\label{fig:grand-diag} Grand potential diagrams for all evaluated systems in the range of oxygen chemical potential corresponding to the annealing process. Green line indicates the cation ratio of the film presented in table \ref{table:dep-parameters}.}
\end{figure*}

The predicted phases and the stability of the ternary compounds have to be further evaluated via the grand potential diagrams to incorporate the effect annealing has on the structure of the mixtures. Figure \ref{fig:grand-diag} presents grand potential diagrams for each of the evaluated systems open to oxygen in the range of oxygen chemical potential corresponding to the annealing process. In this case the diagrams are two-dimensional and depict the energy above hull as a function of the cation ratio, with the ratio determined by XPS represented as a green line. As mentioned before, any compound with its energy above hull higher than zero is classified as unstable by MP, but that classification does not take into account the errors associated with the DFT calculations. Hautier \textit{et. al.} conducted a study on the stability of ternary oxides and found that overall the formation energies calculated by DFT have an accuracy of 24 meV/atom \cite{Hautier2012a}. This becomes particularly important when considering the grand potential diagrams, as a phase could be classified as unstable by MP but have an energy above hull within the accuracy of the DFT calculations. Therefore, we identified three different zones depicting the stability of the phases in the grand potential diagrams: stable, with energies above hull below -24 meV/atom; possibly stable, with energies between -24 meV/atom and 24 meV/atom; and unstable, with energies above 24 meV/atom. This allows one to account for the errors in the DFT calculations when evaluating phase stability for a more accurate assessment.

\onecolumngrid

\twocolumngrid

The grand potentials in figure \ref{fig:grand-diag} are again presented in two panels corresponding to the phase diagrams at 0 K shown in figure \ref{fig:phase-diag}. The top panel presents the systems with no predicted ternary compounds. In all these systems, Zr-Ta-O, Hf-Ta-O and Si-Ta-O, the only phases predicted as possibly stable are Ta$_2$O$_5$ and the dopant oxide. All other phases have energies above hull higher than 5 eV and thus are not presented in the graph. The bottom panel of figure \ref{fig:grand-diag} corresponds to the systems with predicted stable and/or unstable ternary phases. For the systems with predicted unstable ternary compounds, Nb-Ta-O and Ti-Ta-O, the grand potential diagrams show that the only possibly stable phases are Ta$_2$O$_5$ and the dopant oxide. In particular, for Ti-Ta-O the unstable ternary compounds have an energy above hull higher than 0.5 eV, well above the accuracy of the DFT calculations. Lastly, there are three systems for which the phase diagrams showed the presence of at least one stable or possibly stable ternary phase: Sc-Ta-O, Zn-Ta-O and Al-Ta-O. For Sc-Ta-O the most stable phase is the ternary compound ScTaO$_4$ and Ta$_2$O$_5$ and Sc$_2$O$_3$ are also predicted as possibly stable phases. The grand potential diagram for the Zn-Ta-O system shows the ternary compounds Ta$_2$Zn$_3$O$_8$ and Ta$_2$ZnO$_6$ as possibly stable as well as featuring the binaries Ta$_2$O$_5$ and ZnO. Note that the ternary compound Ta$_2$ZnO$_6$ is not featured as part of the convex hull according to the MP database but considering the error associated with the calculations this phase is classified as possibly stable given that its energy above hull is around 0.019 eV. Lastly, for Ta-Al-O, the ternary compound TaAlO$_4$ is predicted to be the most stable followed by Ta$_2$O$_5$ and Al$_2$O$_3$ which are possibly stable. We have disregarded an aluminum oxide from the MP database, Al$_{11}$O$_{18}$, given that it has not been experimentally synthesized. For all the predicted ternary compounds, including Ta$_2$ZnO$_6$, there is no change in the oxidation state of the cations compared to the stable binary oxide phases and thus the reaction necessary to form these compounds involves only the binary oxides (that is, Ta$_2$O$_5$ and the dopant oxide).

\begin{figure}[h!]
	\includegraphics[width=\linewidth]{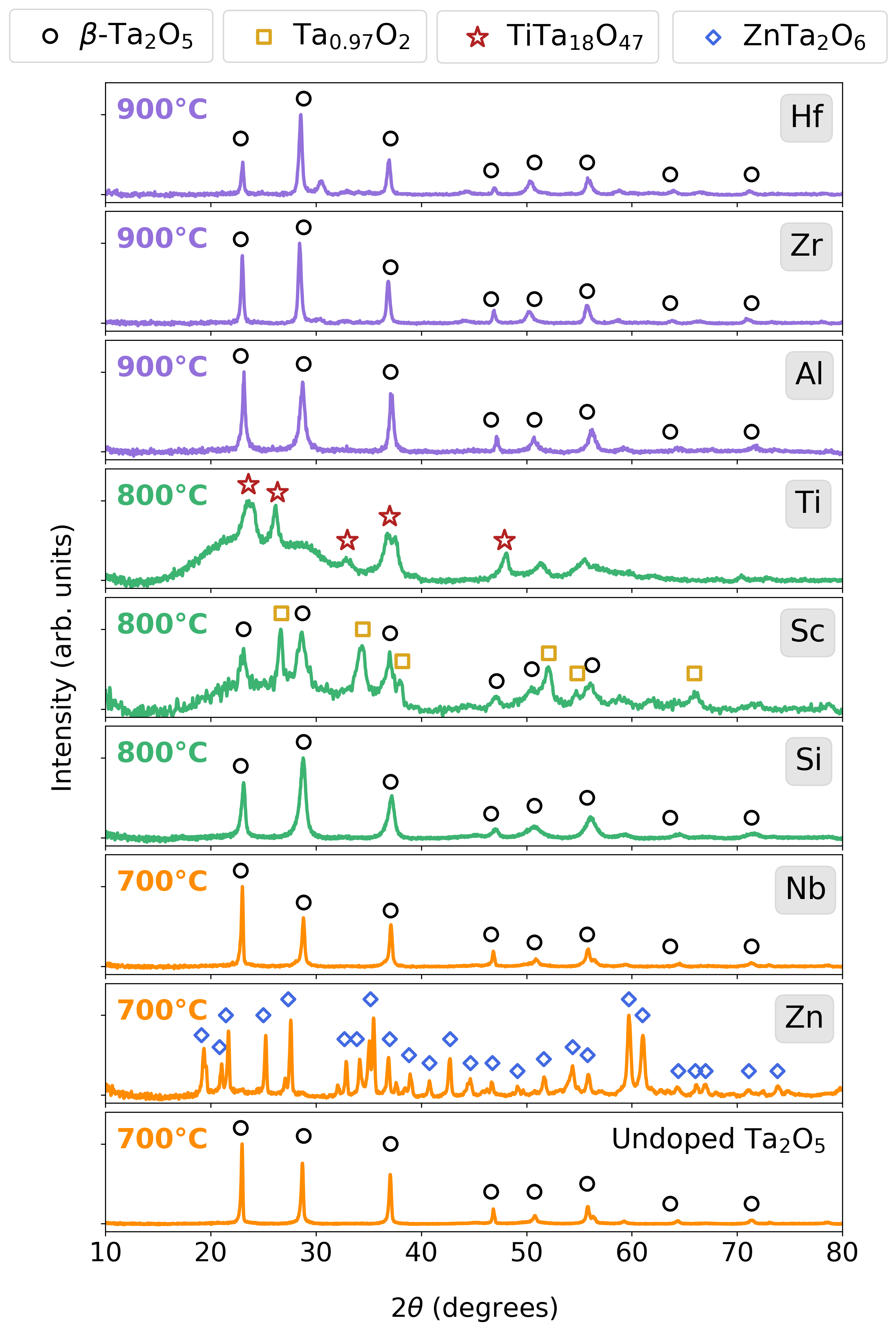}
	\caption{\label{fig:xrd} Diffractograms for all the evaluated films after crystallization. The annealing temperature is indicated in the top left and the dopant cation in the top right of each diffractogram. Tabulated peak positions for $\beta$-Ta$_2$O$_5$ (reference pattern PDF 00-025-0922 \cite{gates2019powder}), Ta$_{0.97}$O$_2$ (reference pattern PDF 00-037-0117 \cite{gates2019powder}), TiTa$_{18}$O$_{47}$ (reference pattern PDF 00-021-1423 \cite{gates2019powder}) and ZnTa$_2$O$_6$ (reference pattern PDF 00-049-0746 \cite{gates2019powder}) are included.}
\end{figure}

We compare these predictions with the structure of the thin films determined by GIXRD. Measurements were performed for all samples as deposited and after each annealing temperature. All as-deposited coatings were amorphous and crystallized after high temperature annealing. Figure \ref{fig:xrd} presents the diffractograms for all the crystallized films. The crystallization temperature is shown in the upper left corner of each diffractogram. Undoped Ta$_2$O$_5$ crystallizes after annealing at 700\C{} in the $\beta$ phase, the low temperature orthorrombic polymorph (reference pattern PDF 00-025-0922 \cite{gates2019powder}).

From the phase and grand potential diagrams shown in figures \ref{fig:phase-diag} and \ref{fig:grand-diag} respectively, the systems predicted to have no ternary phase formation even after undergoing the annealing process correspond to ZrO$_2$-, HfO$_2$-, SiO$_2$-, Nb$_2$O$_5$- and TiO$_2$-doped Ta$_2$O$_5$. For all these systems, the only predicted possibly stable phases are Ta$_2$O$_5$ and the dopant oxide. As observed in figure \ref{fig:xrd}, the crystallized phase for ZrO$_2$-, HfO$_2$-, SiO$_2$- and Nb$_2$O$_5$-doped Ta$_2$O$_5$ is Ta$_2$O$_5$ in its $\beta$ phase. The amorphous phase corresponds to the dopant oxide which is verified by XPS measurements that show the oxidation state of the dopants is consistent with their corresponding binary oxide. The fact that the Ta$_2$O$_5$ phase crystallizes first is to be expected given that it is the most abundant phase. The low dopant concentration in the films possibly hinders the nucleation of the dopant oxide phase necessary for its crystallization. In the case of TiO$_2$-doped Ta$_2$O$_5$, the prediction fails because the film crystallizes in the ternary compound TiTa$_{18}$O$_{47}$ (reference pattern PDF 00-021-1423 \cite{gates2019powder}) as shown previously in \cite{fazio2020structure}. However, this ternary compound is not included in the MP database and therefore is not featured in the phase diagram. This exemplifies that the genomic approach the MP database provides has limitations.


Sc$_2$O$_3$- and Al$_2$O$_3$-doped Ta$_2$O$_5$ both have ternary phases, ScTaO$_4$ and TaAlO$_4$ respectively, that are predicted to be the most stable while Ta$_2$O$_5$ and the dopant oxide are also expected to be possibly stable. However, the crystallized films do no feature any ternary compounds. In the case of Sc$_2$O$_3$-doped Ta$_2$O$_5$, the addition of Sc$_2$O$_3$ induces the crystallization of Ta$_2$O$_5$ and a tantalum suboxide (Ta$_{0.97}$O$_2$, reference pattern PDF 00-037-0117 \cite{gates2019powder}) due to oxygen competition, as discussed in \cite{fazio2020growth}. This competition is brought by the fact that these oxides have similar Gibbs free energy of formation and thus similar formation enthalpies. However, based on table \ref{table:thermo-prop}, the calculated formation enthalpy of Ta$_2$O$_5$ is 18\% lower than that of Sc$_2$O$_3$ when experimentally they only differ by less than 7\% at 298 K. The fact that the difference in formation enthalpy between Ta$_2$O$_5$ and Sc$_2$O$_3$ is higher for the calculations at 0 K than for the experimental values at 298 K could result in an absence of oxygen competition effects in the calculations, while experimentally those effects are even observed in the amorphous as-deposited film \cite{fazio2020structure}. For Al$_2$O$_3$-doped Ta$_2$O$_5$, the film crystallizes as $\beta$-Ta$_2$O$_5$ while the dopant oxide phase remains amorphous. These two phases are predicted to be possibly stable, but the most stable according to the calculations is TaAlO$_4$. The absence of this ternary compound in the crystallized coating could be attributed to the fact that this phase has a cation ratio much larger than the cation ratio of the film. Significant Al cation diffusion would be necessary to form the ternary compound, as the dopant seems to be distributed homogeneously in the film given that XPS measurements do not detect significant variations in atomic concentrations for different regions in the coating surface.

The grand potential diagram for ZnO-doped Ta$_2$O$_5$ in figure \ref{fig:grand-diag} indicates that there are two possibly stable ternary compounds: Ta$_2$Zn$_3$O$_8$ and Ta$_2$ZnO$_6$. Ta$_2$O$_5$ and ZnO are also predicted to be possibly stable phases. The film crystallizes as the ternary compound, Ta$_2$ZnO$_6$ (reference pattern PDF 00-049-0746 \cite{gates2019powder}), as shown in figure \ref{fig:xrd}, instead of the ternary Ta$_2$Zn$_3$O$_8$ or even the binary oxides which compose the convex hull. Once again, the ternary phase with a cation ratio being higher than the cation ratio of the film is not stabilized in the crystallized coating. In this case, Ta$_2$Zn$_3$O$_8$ has a cation ratio of 0.6 which is three times higher than the film cation ratio while Ta$_2$ZnO$_6$ has a cation ratio of 0.3, close to the ratio of the film.

\begin{figure}[h!]
	\includegraphics[width=0.8\linewidth]{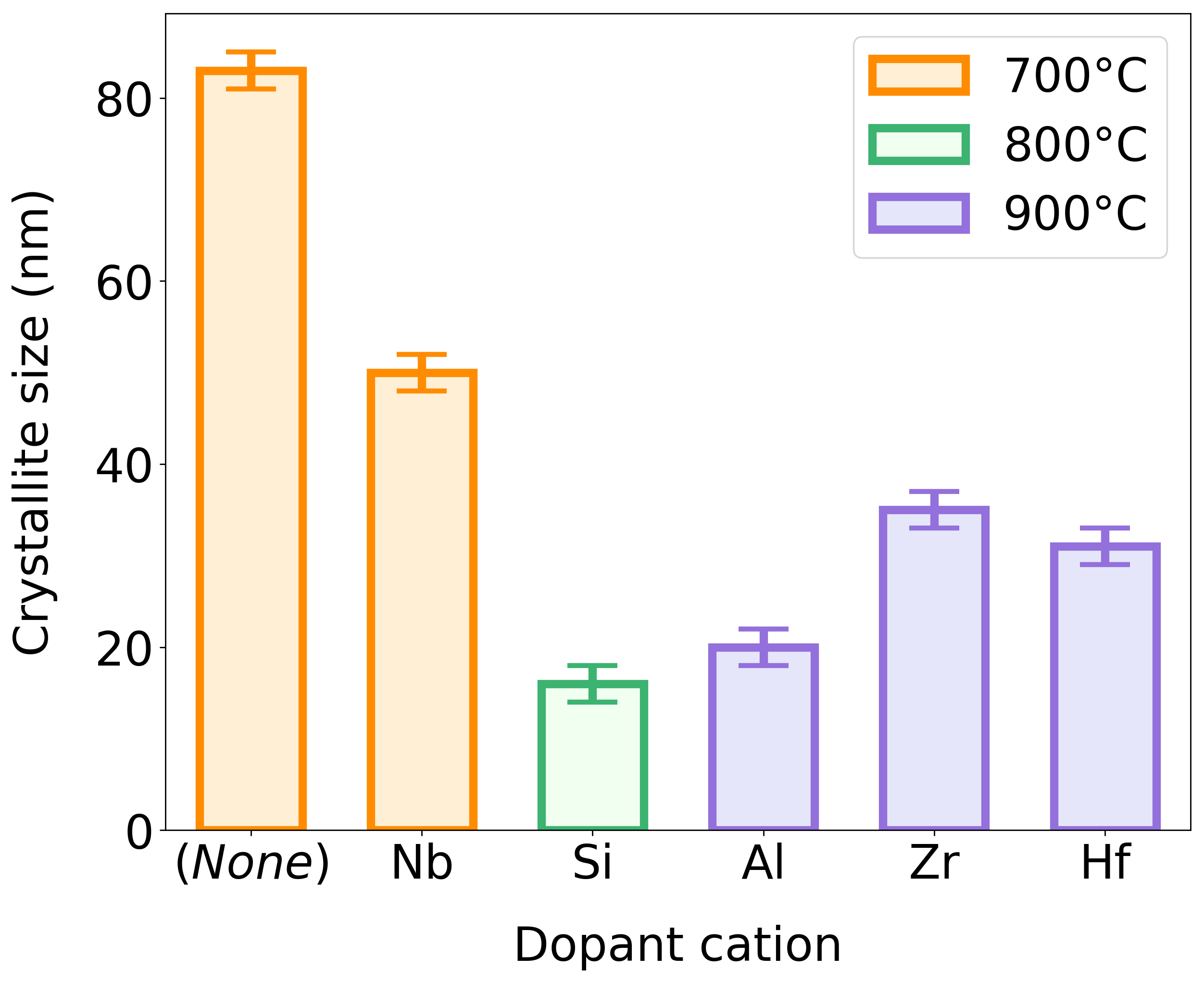}
	\caption{\label{fig:crystallite} Ta$_2$O$_5$ crystallite size estimated by the Scherrer equation for an undoped film and for different mixtures. The corresponding crystallization temperature for each coating is indicated. Errors were estimated from the uncertainties in peak width and position.}
\end{figure}

For the mixtures in which only Ta$_2$O$_5$ crystallizes, figure \ref{fig:crystallite} presents the Ta$_2$O$_5$ crystallite size estimated applying the Scherrer equation \cite{langford1978scherrer} to the (001) peak at $2\theta \approx 22.83^{\circ}$. All the mixtures have smaller crystallite size than the undoped film which indicates that these dopants suppress the crystallization of Ta$_2$O$_5$. SiO$_2$-doped Ta$_2$O$_5$ presents the largest reduction in crystallite size of all the evaluated coatings but does not have the highest crystallization temperature. Al$_2$O$_3$-, ZrO$_2$- and HfO$_2$-doped Ta$_2$O$_5$ crystallize at the highest temperature with crystallite sizes from 20 nm to 35 nm. In the case of Nb$_2$O$_5$-doped Ta$_2$O$_5$, the crystallite size is reduced by 40\% but there is no apparent increase in the crystallization temperature given the 100\C{} annealing steps used in this study. Figure \ref{fig:crystallite} also shows that higher crystallization temperatures do not correlate to smaller crystallite sizes, which could be due to the fact that the dopant cation concentration is not strictly comparable across these mixtures ranging from around 0.12 for Nb to 0.26 for SiO$_2$. Tewg \textit{et. al.} studied ZrO$_2$-doped Ta$_2$O$_5$ coatings with different cation concentrations \cite{tewg2004suppression} and found that increasing the cation ratio from 0.19 to 0.33 reduces the Ta$_2$O$_5$ crystallite size by 40\% which is correlated with an increase of the crystallization temperature from 800\C{} to 900\C{}. Overall, the observed increase of crystallization temperature and reduction of crystallite size in these mixtures indicate that Al$_2$O$_3$, SiO$_2$, ZrO$_2$, Nb$_2$O$_5$ and HfO$_2$ act as amorphizer agents inducing structural disorder into the amorphous phase and thus increasing the thermal stability of the amorphous coatings. The least effective amorphizer agent is Nb$_2$O$_5$, which is the only one that has similar characteristics to Ta$_2$O$_5$ such as the oxidation state of the cation and the ionic radius. For the other dopants, their effectiveness might still be further improved by varying the cation concentration.


\section{Conclusions} \label{sec:conclusions}

Phase diagrams at 0 K based on DFT calculations were constructed for different Ta$_2$O$_5$-based mixed oxide systems using the MP database. In order to evaluate the phase stability taking into account the thermodynamics of the annealing process to which the coatings are subjected to, grand potential diagrams were also calculated for the corresponding oxygen chemical potential values. The predictions of phase stability provided by these diagrams were in good agreement with the experimentally measured crystallized phases of the mixed amorphous oxide films. For systems in which no ternary compounds are formed, such as SiO$_2$-, ZrO$_2$-, Nb$_2$O$_5$- and HfO$_2$-doped Ta$_2$O$_5$, both Ta$_2$O$_5$ and the dopant oxide are predicted to be stable however given the low cation ratio of the films only Ta$_2$O$_5$ crystallizes. In all these systems, the dopant acts as an amorphizer agent increasing the thermal stability of the amorphous coatings. It was found that Nb$_2$O$_5$, the compound most chemically similar to Ta$_2$O$_5$, is the least efficient amorphizer agent. For the cases in which a ternary compound was predicted to be stable, those phases with cation ratios much larger than the cation ratio of the film cannot be stabilized. This finding shows that certain kinetics considerations, such as cation diffusion necessary to form these compounds, are not captured by the DFT calculations. It was found that the MP database did not contain a ternary compound found experimentally in TiO$_2$-doped Ta$_2$O$_5$ leading to incorrect predictions. Also, DFT overestimated the formation enthalpy difference between Sc$_2$O$_3$ and Ta$_2$O$_5$ and did not reproduce the oxygen competition effects observed in those coatings. Overall, the good agreement found between the calculations and the experimental results aided in identifying potential amorphizer agents for Ta$_2$O$_5$ and predicting the presence of ternary compounds for the specific annealing conditions used in the experiments. Given the large available databases of DFT calculations it might be possible to now apply a genomic approach for \textit{in silico} design of mixed oxide amorphous coatings for improved thermal stability or for ternary oxide compound formation. This analysis can lead the way to applying the powerful genomic approach to a new range of applications involving amorphous materials such as optical coatings, energy storage and electronic devices with tailored properties.

\begin{acknowledgments}

This work was supported by the National Science Foundation - Moore Foundation Center for Coatings Research under NSF award No.	1708010 and NSF award No. 1710957. The authors would like to thank Kiran Prasai for his helpful input and comments.
\end{acknowledgments}

\bibliography{survey_paper}

\end{document}